\documentstyle[aps,psfig,twocolumn,pre]{revtex}

\def\fw{\columnwidth}

\begin{document}

%\draft
\title{A lattice gas model of avalanches in a granular pile}

\author{Antal K\'arolyi$^{1,2}$ and J\'anos Kert\'esz$^1$ }

\address{$^1$ Department of Theoretical Physics, Technical University of 
Budapest, Budafoki u. 8, 1111 Hungary\\
$^2$ Theoretische Physik, FB 10, Gerhard-Mercator Universit\"at, 47048
Duisburg, Germany}

\date{\today}

\maketitle

\begin{abstract}

A granular media lattice gas (GMLG) model is used to study avalanches in a
two-dimensional granular pile.  We demonstrate the efficiency of the
algorithm by showing that several features of the non-critical behaviour of
real sandpile surfaces, such as the bounded outflow statistics or the
finite-size effect of the time evolution of the pile mass, can be
reproduced by this simulation approach.

\end{abstract}


\bigskip

%\begin{multicols}{2}

The concept of self-organized criticality (SOC) \cite{Bak} has become
popular in the research of non-equilibrium systems.  According to this
idea, weakly driven non-equilibrium systems can spontaneously organize
themselves into a state of diverging characteristic length and time
scales.  It was suggested that avalanches on sandpile surfaces may fit into
the SOC framework and this idea inspired considerable activity in the field
of the physics of granular materials.

Sandpiles soon proved to behave differently from the proposed picture in
many aspects. Avalanches in a rotating drum were found to occur at
well-defined interwalls and the probability density of avalanche durations
turned out to be sharply peaked \cite{Jaeger1}. Using a different
experimental setup, where grains were dropped onto the top of a conical
pile, Held {\it et al.}\  did find scaling in small piles\cite{Held}.
However, if the size of the pile was above a certain value, the small
fluctuations disappeared from the system. The crossover to this
quasi-regular behaviour can be explained by the fact, that the surface of
the pile has two characteristic angles rather than a single one: the angle of
repose and the angle of marginal stability. In the case of sufficiently
small piles the size of one single particle is large enough to raise the
local angle above the critical value and therefore no hysteresis can be
found\cite{Liu}. Further measurements pointed out that the intervals
between two consecutive large events contain several small avalanches with
a power law size distribution\cite{Bretz}.  There have been attempts to
predict the large events, but large avalanches appear to show the
characteristics of a Markov-process \cite{Morales}.  In accordance with
this result, the power spectrum computed from the time sequence of
avalanches follows a $1/f^2$ rule (see e.g.  \cite{Held}). By
re-analyzing the results of several experimental investigations, J.
Feder\cite{Feder} has found that in the case of the small events stretched
exponential functions (having a characteristic size) fit the avalanche size
distributions more precisely.

Recently two model systems have been found which do show SOC: in rice
piles (experimental results: \cite{Frette}, simulation:
\cite{Amaral}) and avalanches of small particles in a bidisperse
filling of a rotating drum\cite{Baumann} (simulation). In both cases it
turns out that criticality is a consequence of surface roughness, due to
the elongated grains in the former case and the surface niches formed by
the larger particles in the latter.  It is also surprising that rice piles
made up of grains of different aspect ratio - so varying only in a material
constant in terms of critical behaviour - belong to different universality
classes\cite{Amaral}.

The aim of this work is to present the Granular Media Lattice Gas (GMLG)
model through a study of avalanches in a pile.  Experimental results often
lack large enough statistics, molecular dynamics is inadequate for the same
reason, while simplified computer models may exclude important aspects of
real sandpile surfaces (such as the difference between the two critical
angles or inertia effects).  The simulation method we use here is a
promising compromise between the two: It has been designed to describe many
features of real sandpiles while it is based on a fast parallel algorithm.

The GMLG model is the generalization of a successful, fully discrete
hydrodynamic algorithm\cite{Frisch,lgpardif}. It is defined on a triangular
lattice, where indistinguishable point-like particles either travel at unit
velocity along the lattice bonds or they can be at rest at the nodes and on
the bonds. An update consists of the collision and propagation step.
During the collision step the particles are scattered at the lattices nodes
while during propagation step the moving particles are transferred to the
nearest neighbour sites.

While bulk collisions of the original hydrodynamic model conserve mass,
energy and momentum, in the GMLG model energy dissipation and friction
effects are also taken into account.  As a result of the restricted set of
velocities (0 or 1), material parameters can be introduced only in a
stochastic way, by means of probability variables.

A driven sandpile is an example of such a granular system, where both
fluidized and static regions can be present at the same time.  The
different behaviour of the material within these regions is reflected in the
collision rules of the model as follows. 

\begin{figure}
\centerline{\psfig{width=\fw,figure=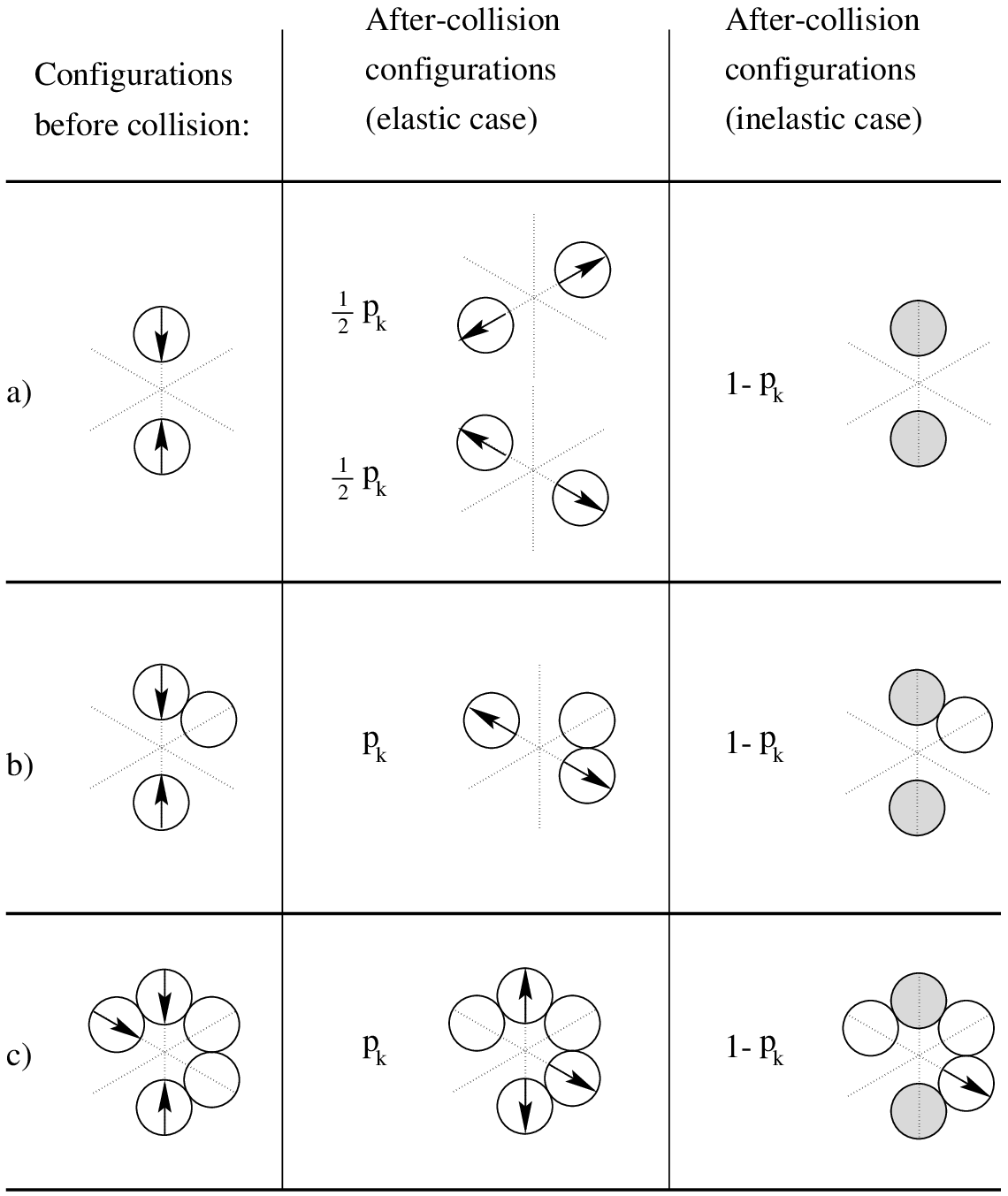}}
\caption{}
\label{ncoll}
\end{figure}

In fluidized regions momentum is conserved, but collisions can dissipate
energy. This is described by the parameter $p_k$, which is the model
equivalent of the energy restitution coefficient.  If energy is conserved
at a particular site - with a probability of $p_k$ - then the collision
rules are similar to that of the hydrodynamic model, with the exception,
that rest particles may block possible scattering directions. If this is the
case, we choose the after-collision configuration which provides the maximum
mixing of states. If the collision is dissipative, then the maximum energy
is dissipated while conserving momentum.  We present three examples on
Fig. \ref{ncoll} demonstrating how these principles work in the fluidized
region.

\begin{figure}
\centerline{\psfig{width=5 truecm ,figure=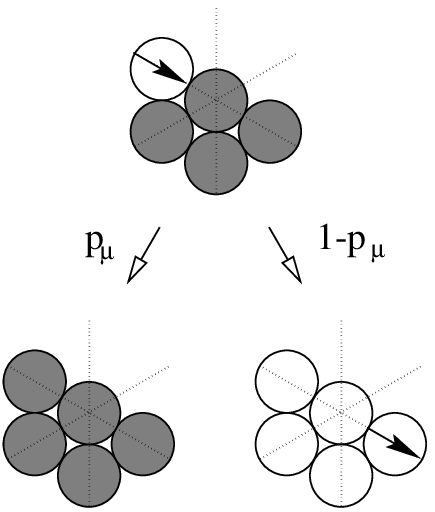}}
\caption{}
\label{nfric}
\end{figure}

The compact static part of the pile behaves like a solid with a large mass,
where friction effects are to be taken into account. When moving particles
interact with the bulk, their momentum can be transferred through the force
chains to the walls of the vessel. This is modelled by the friction
variable $p_\mu$. It gives the probability, with which a moving particle
stops when arriving at a bulk site (see Fig. \ref{nfric}). Bulk particles are
such rest particles which are supported either by another bulk site or the
bottom of the vessel. Although this definition itself does not guarantee
that there cannot be nodes that temporarily are not supported by the bottom
or the walls -- in principle, it would be possible to set up an algorithm
that finds the bulk sites at the cost of efficiency -- but in case of
simple geometries our definition works correctly. (This process is similar
to the {\it capture} process, introduced in a different
context\cite{Hernan}.)

The implementation of gravity for moving particles is straightforward,
like for the hydrodynamic models\cite{lgpardif}.  Gravity rules applied to
rest particles involve an effective metastability, in that moving particles
can set off rest ones. This metastability is responsible for the hysteresis
of the pile surface.  Note, that in our case the ``microscopic'' rules
result in the two different critical angles, we do not include them, what
is the usual approach with cellular automaton sandpile models (e.g.
\cite{Bak,Amaral,Prado}).

\begin{figure}
\centerline{\psfig{width=\fw,figure=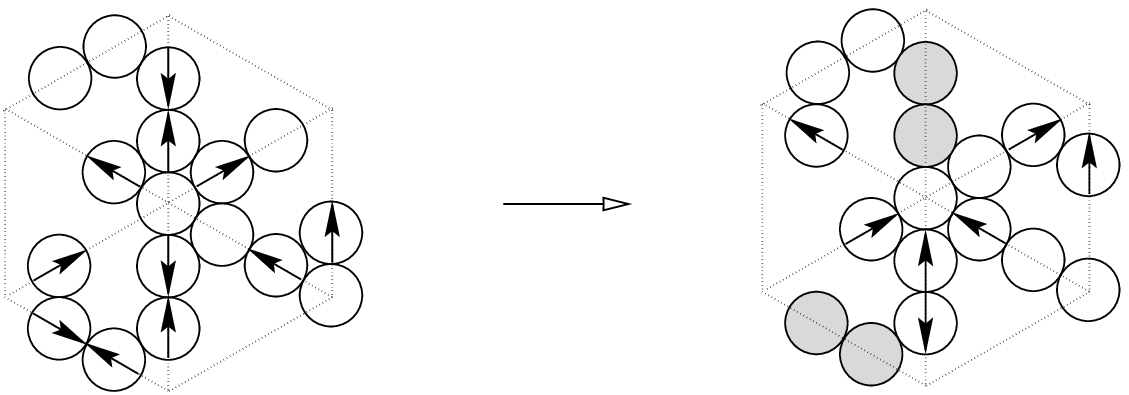}}
\caption{}
\label{nprop}
\end{figure}

After applying the collision, friction and gravity rules, the particles
are transferred to the nearest neighbour sites. The GMLG propagation step
differs from the hydrodynamic model in one aspect. Dissipative binary
collisions may take place here, since there can be up to two particles on
each bond between NN sites, as it is illustrated on Fig. \ref{nprop}.

With the model described above and with similar approaches it is possible
to simulate such different systems as pipe flows, shaken boxes, granular
mixtures or static
piles\cite{KK,Herrmann1,Herrmann2,Herrmann3}. In this paper we examine the
properties of avalanches.

\begin{figure}
\centerline{\psfig{width=\fw,figure=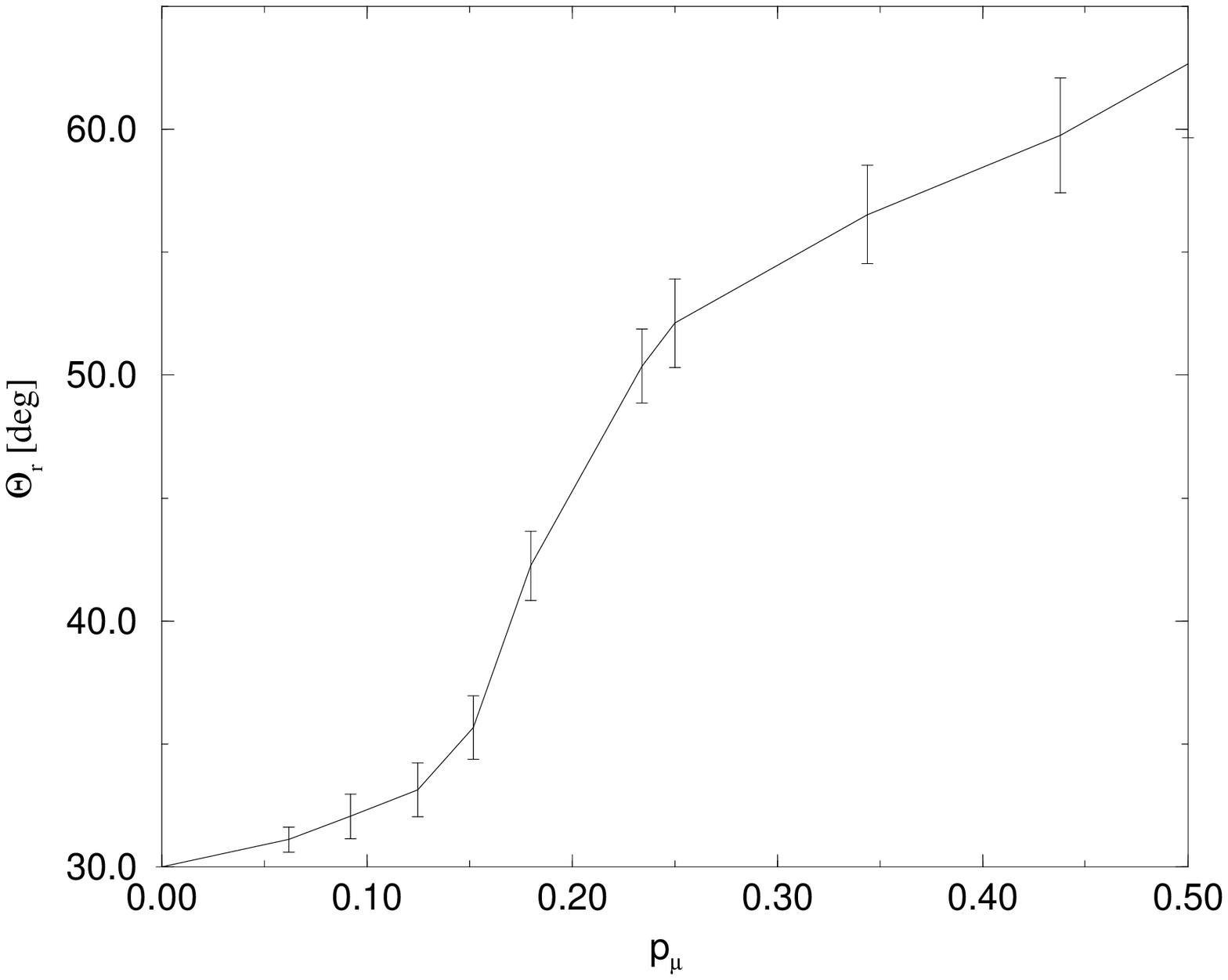}}
\caption{}
\label{calibrate}
\end{figure}

As a first test of the model we calculate the dependence of the angle of
repose of the pile on the friction variable $p_\mu$.  We have chosen a
method described in \cite{Boutreux}, the steady-state filling of a silo and
the result is shown on Fig. \ref{calibrate}. The error bars give information
about the difference between the angle of repose and the angle of marginal
stability.  The angles are rather high compared to real-world systems,
which is a consequence of the underlying lattice (the smallest angle of
repose is $30^o$ in the model). 

\begin{figure}
\centerline{\psfig{width=5truecm,figure=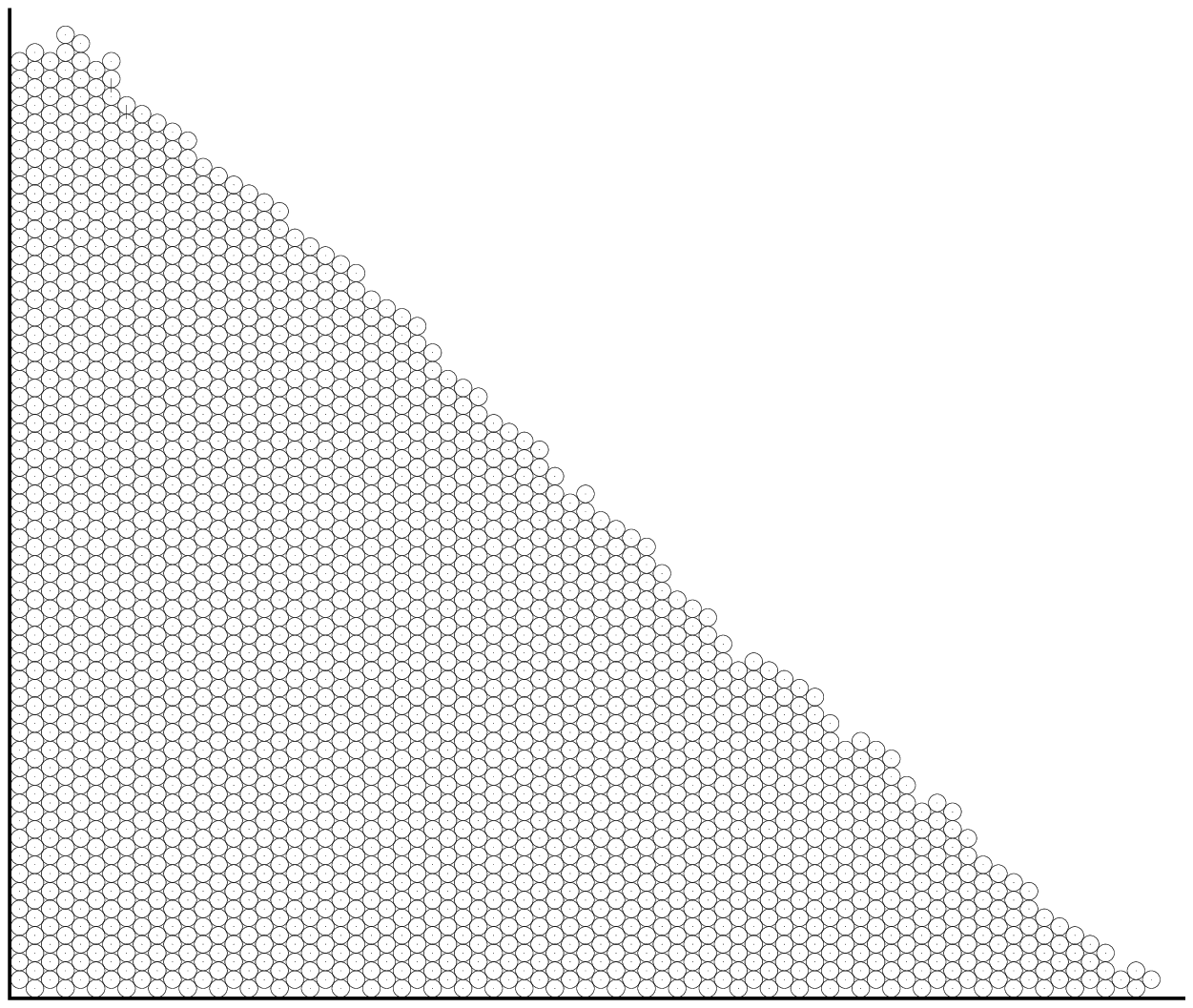}}
\caption{}
\label{halfpile}
\end{figure}

The system studied - a 2D box with one open side - is a classical setup for
examining avalanches and is also analogous with the Hele-Shaw cell used e.g.
by Frette {\it et al}\cite{Frette}. The particles are dropped near to the
side-wall in such a way that a half-pile is building up (see
Fig. \ref{halfpile}). The pile is driven quasi-statically, that is the
particles are added one by one, after all activity caused by dropping the
previous particle ceased. The size $L$ of the system is defined by the
length of the horizontal support. The measurements start after the
stationary state has been reached.

\begin{figure}
\centerline{\psfig{width=\fw,figure=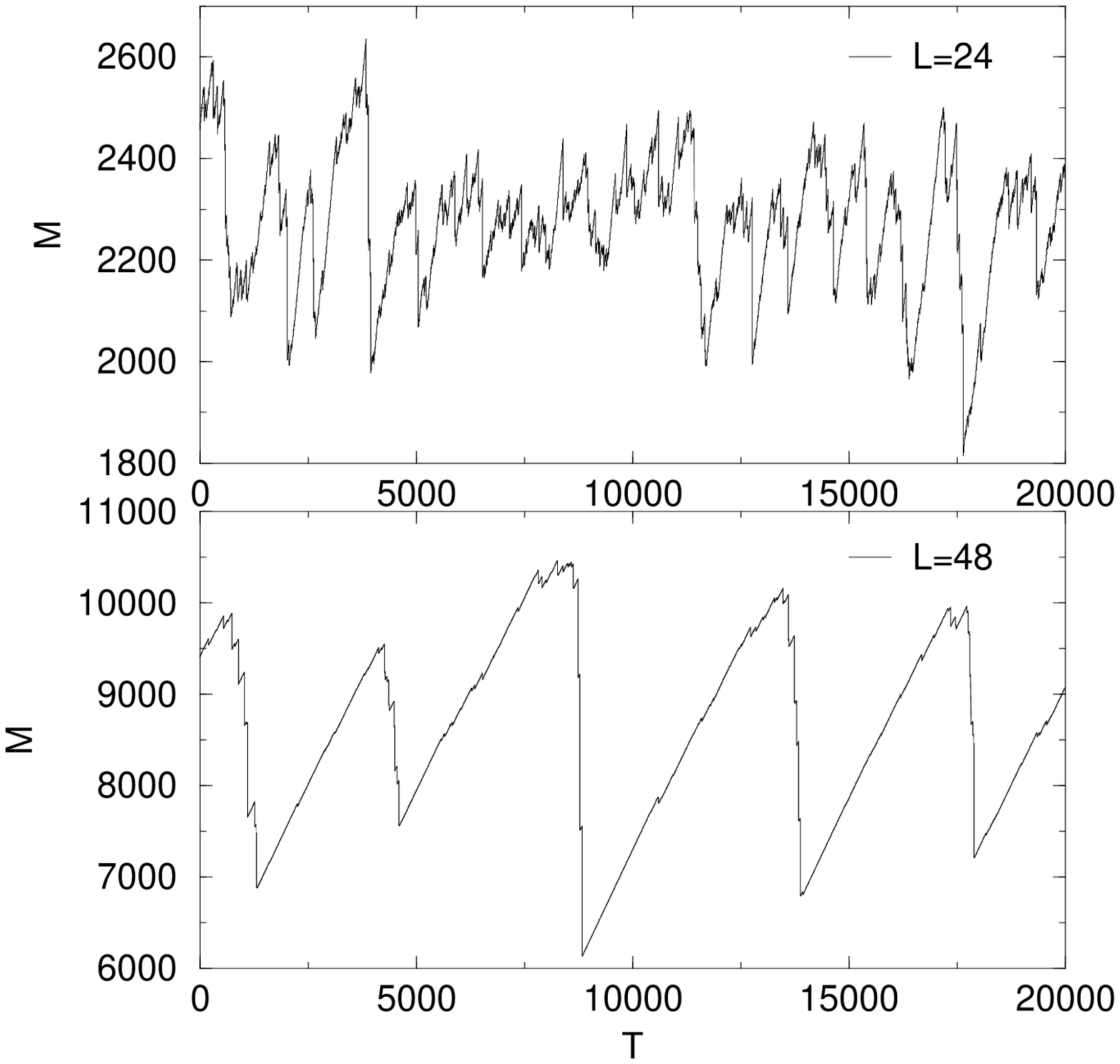}}
\caption{}
\label{tdep}
\end{figure}

First we consider the time evolution of the total mass of the pile (see
Fig. \ref{tdep}). The time unit here is the interval between dropping two
consecutive grains, which can last from one single update to several
hundred updates long. 

\begin{figure}
\centerline{\psfig{width=\fw,figure=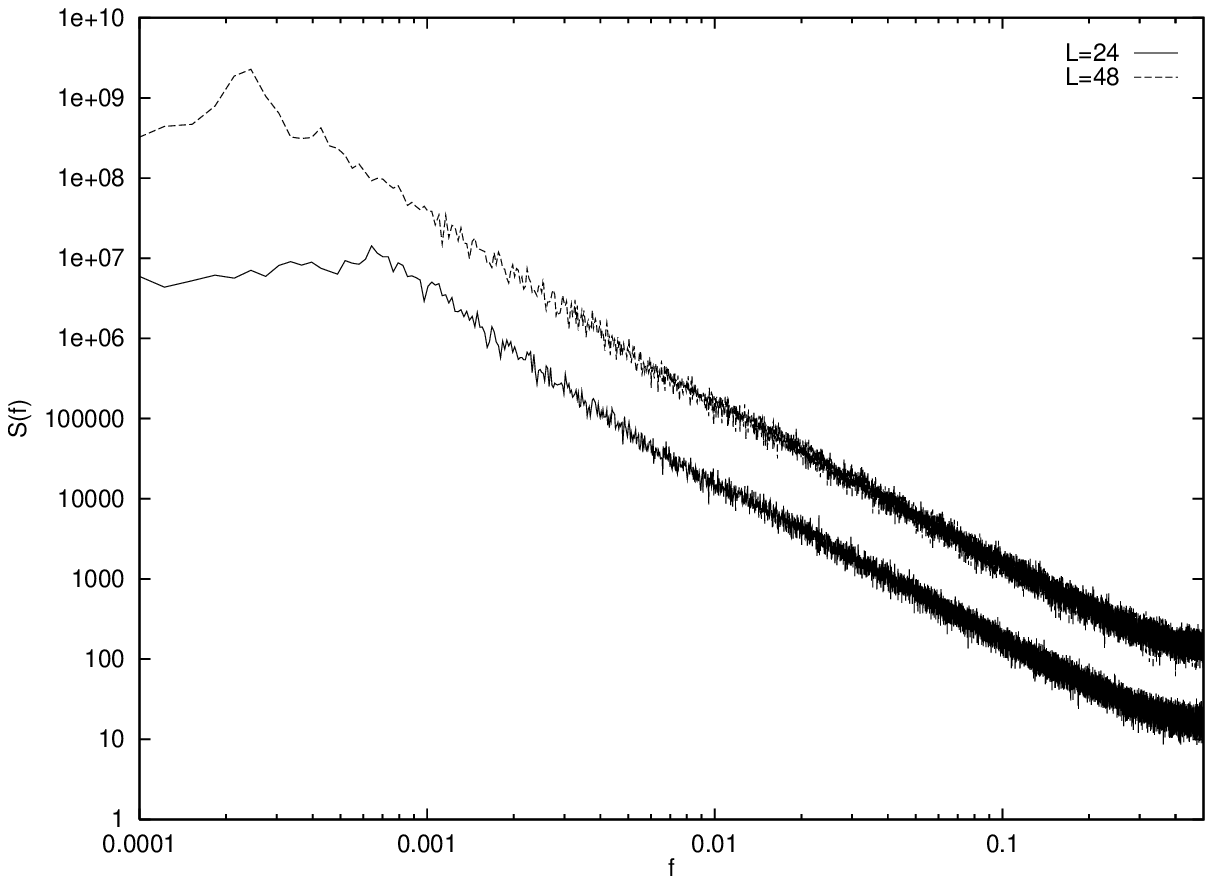}}
\caption{}
\label{power}
\end{figure}

The graphs on Fig. \ref{tdep} represent two runs, where all material parameters
are kept identical, while the system sizes are varied. Although the sizes
are modified only by a factor of two, the curves are qualitatively
different. The first one ($L=24$) is a function irregularly fluctuating on
many time scales, but the second one ($L=48$) is much more regular,
quasi-periodic with a period of about $4000$ timesteps.  This can be
underlined by comparing the power spectra of the two data series (see
Fig. \ref{power}).  In case of the larger pile a peak develops, which
corresponds to a frequency of $1/4000$.  This finite-size effect is in nice
agreement with the experimental findings of Held {\it et al} \cite{Held}.
This result also makes it possible to calibrate the length of the 
model system to experimental scales. 

As a next step we study the distributions of avalanches.  We use two
quantities which are sufficient to characterize the size of an avalanche:
the $T$ lifetime of an event (in update units) and the number of particles
falling off the support, that is the $M$ mass of a droplet.  Note that $M$
does not contain information about the small avalanches not reaching the
rim of the support.  A probability density curve contains data obtained
from typically $10^6$-$10^7$ updates.  In \cite{Manna} it was found for a
SOC automaton that the outflow statistics has multiscaling properties. The
quality of our generated data was not good enough to check for this
property.  Therefore, in a simple finite-size scaling framework we look for
the distributions of these quantities in the following form:

$$ p(T,L)\sim L^{-\alpha} f_T(T/L^{\lambda}) $$

$$ p(M,L)\sim L^{-\beta} f_M(M/L^{\mu}) $$.

\begin{figure}
\centerline{\psfig{width=\fw,figure=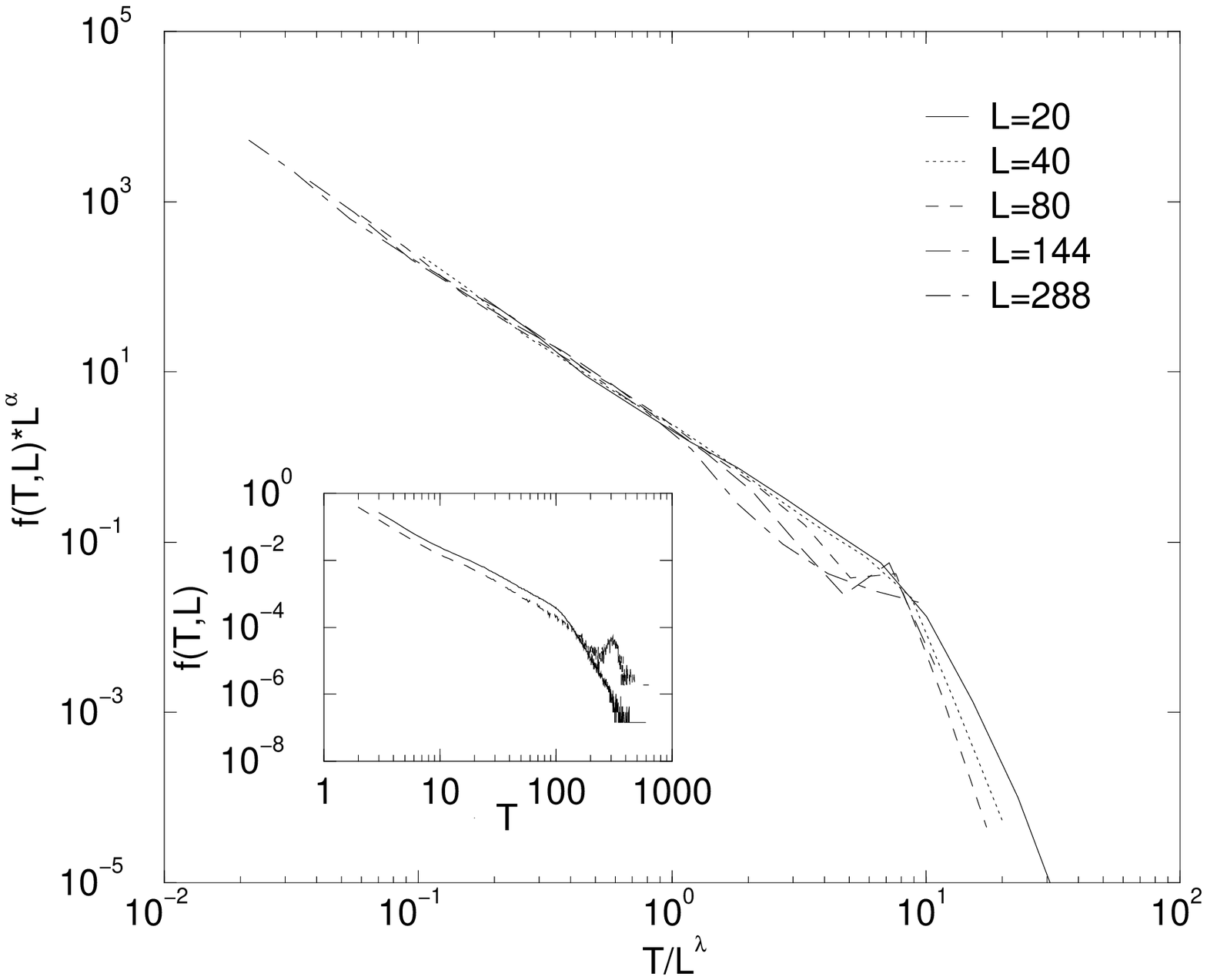}}
\caption{}
\label{tscal} 
\end{figure}

\begin{figure}
\centerline{\psfig{width=\fw,figure=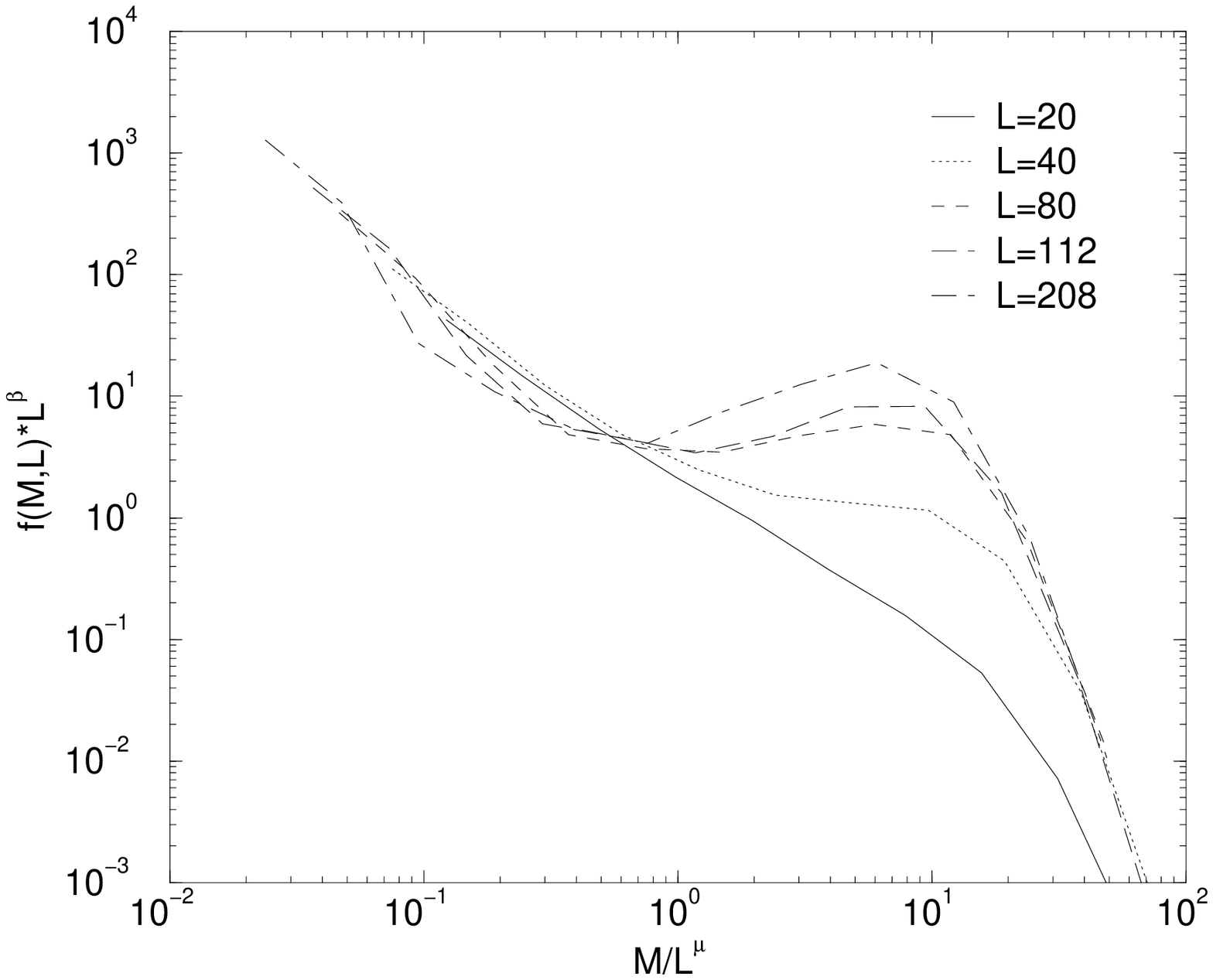}}
\caption{}
\label{mscal}
\end{figure}

Fig. \ref{tscal} and Fig. \ref{mscal} show the probability densities of these
quantities for different system sizes. All curves were logarithmically
binned and rescaled using the ansatz above, but for comparison the inset on
Fig. \ref{tscal} displays two typical raw data curves.

The first apparent feature is that the finite-size effect observed in the
time evolution of the pile appears in the avalanche statistics too.  Both
the lifetime and the droplet probability densities have a power law form
(straight line on a log-log plot) with a sharp cutoff for small system
sizes ($L<40$) while a pronounced peak develops for larger sizes.  This
behaviour was also observed in the IBM experiment \cite{Held}.  This feature
is somewhat less obvious in the statistics of the duration times, since the
sharper characteristic peaks are smoothed when binning the curves, but they
are well-defined, see the original data curves on the inset of
Fig. \ref{tscal}.

The good data collapse demonstrates, that the finite-size scaling ansatz
is well suited for the duration time distributions. A well-defined exponent
can be found for the small avalanches:  $p(T,L) \sim T^y$ with
$y=1.92\pm0.05$. The numerically found exponents are also consistent with
the $y=-\alpha/\lambda$ criterion, which follows from the properties of the
$f_T(a)$ scaling function for $a\to 0$.  Note, that the value of $y$ is
very close to the critical exponent found in a so-called {\it local}
version of a class of rice pile models\cite{Amaral}. The droplet
distributions, however, are different in the two models.

\begin{figure}
\centerline{\psfig{width=\fw,figure=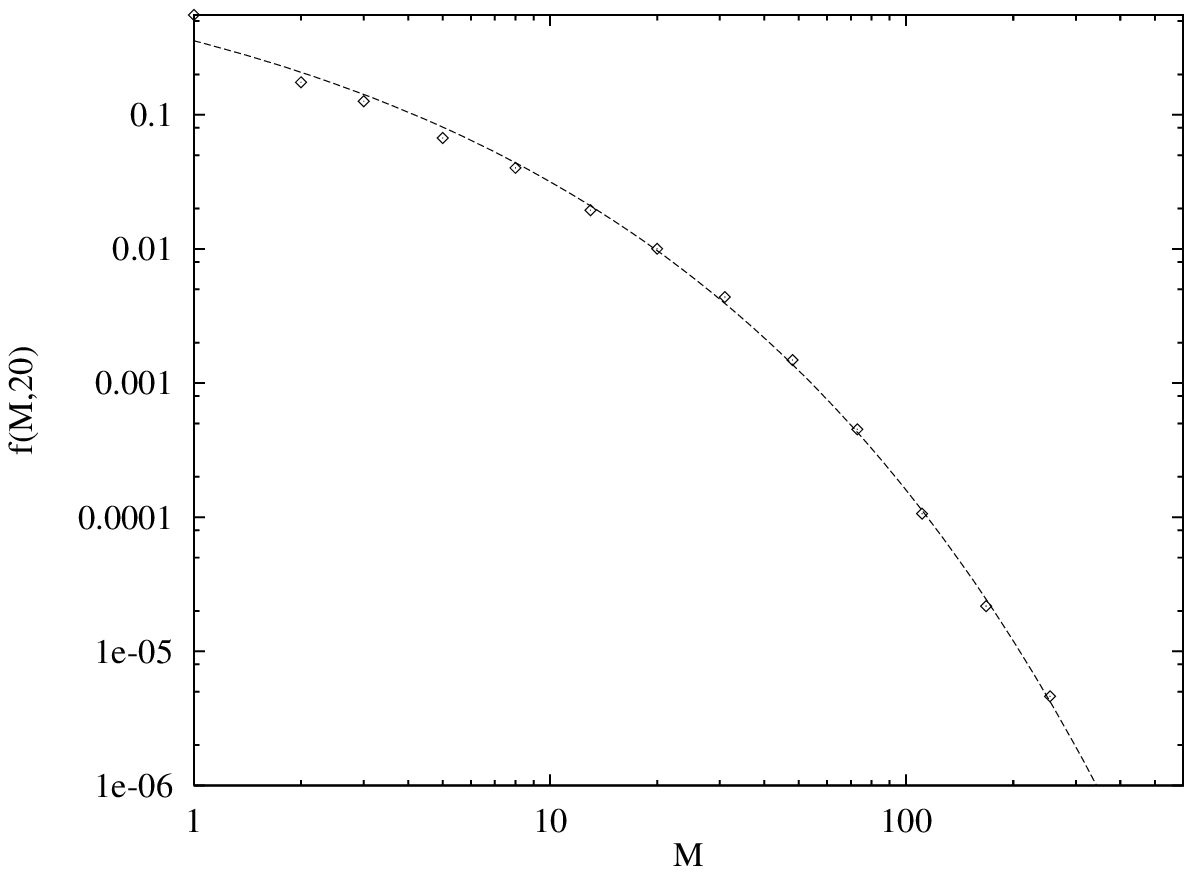}}
\caption{}
\label{strech}
\end{figure}

The outflow statistics of the small, $L=20$ pile needs more attention.
Although the distribution of small droplets can be described by a power
law, it is also possible to fit the whole curve, including large
avalanches, with a stretched exponential ansatz of the form $f(M,20)\sim
\exp(-c*M^\gamma)$. The exponent ($\gamma=0.34$) that was used by
Feder\cite{Feder} while re-analyzing experimental data gives an excellent
fit (see Fig. \ref{strech}).

\begin{figure}
\centerline{\psfig{width=\fw,figure=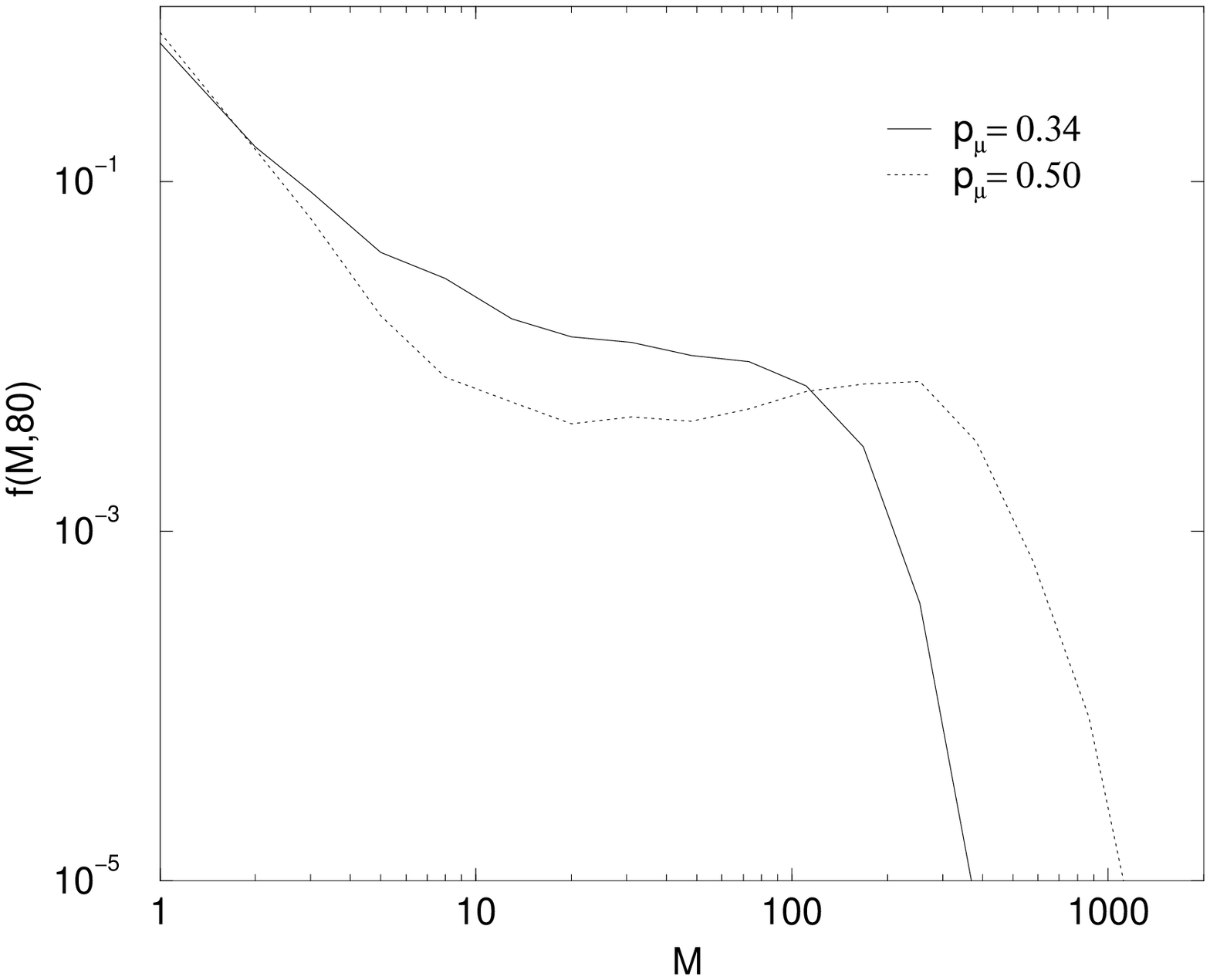}}
\caption{}
\label{fricdep}
\end{figure}

Finally we compare the GMLG model to a recent experiment concerning rice
piles\cite{Frette}. The friction rule in the simulation can be interpreted
as the probability for a grain to get trapped in a local surface minimum,
therefore a higher $p_\mu$ (in terms of the rice pile experiment a higher
$p_\mu$ can be regarded as a higher grain aspect ratio) should result in a
rougher surface and this is what can actually be seen in our model. This
may suggest a smooth transition from the sandpile-type distribution to that
of the rice pile.  By increasing the coefficient of friction, the outflow
probability distribution does get broader, but there is no change in the
scaling properties for different $p_\mu$ values. Higher $p_\mu$ means a
larger angle of repose and also a larger difference between the critical
angles (Fig. \ref{calibrate}), therefore the big avalanches consist of more
particles. Fig. \ref{fricdep} shows outflow distributions for two different
friction values.

\begin{figure}
\centerline{\psfig{width=\fw,figure=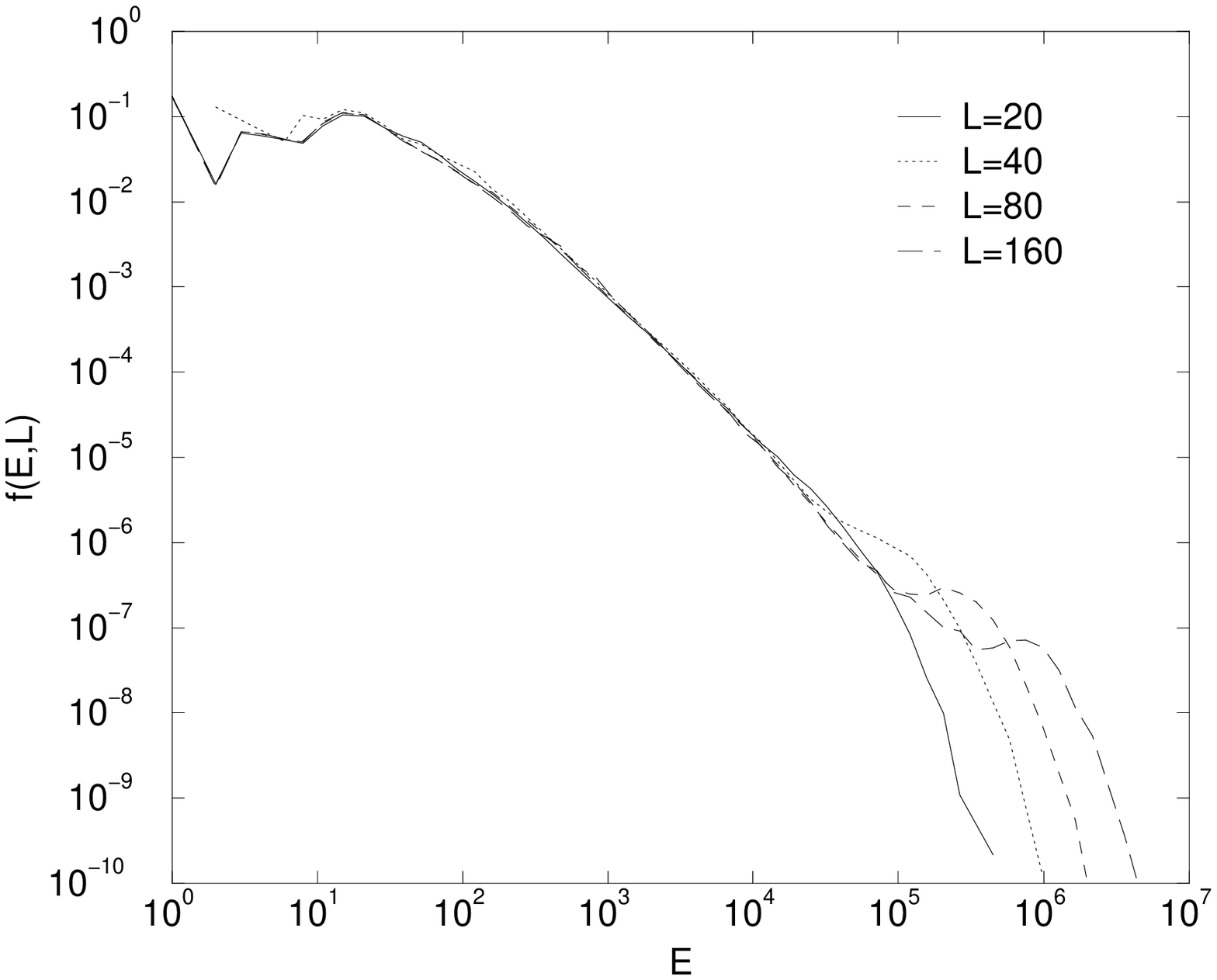}}
\caption{}
\label{dissip}
\end{figure}

Although the droplet and lifetime distributions are sufficient to describe
the avalanche statistics, for the sake of an easier comparison we have
also checked the distribution of the dissipated energy in an avalanche. By
comparing the total potential energy of the pile before and after an
avalanche the dissipated energy can be obtained. The distributions
for four system sizes are plotted on Fig. \ref{dissip}. The finite-size
effects found in lifetime- and droplet distributions can be observed here,
too. The power law part of the probability densities have an exponent of
$1.63\pm0.06$, which is different from the one found experimentally for
rice piles ($\approx 2.0$). The model exponent is very close to the value
measured in an another cellular automaton rice-pile model\cite{Amaral}.

In summary, the GMLG-pile has many features in common with real sandpiles.
The non-trivial finite-size effects, such as the characteristically
different time dependence of the pile mass at different sizes, the outflow
statistics are in good qualitative agreement with experimental findings.
With a proper calibration of the geometry the model may even give
quantitatively good results. It is important to emphasize that the
GMLG-model does not involve any built-in assumptions about the nature of
avalanches (as opposed to many cellular automaton models).

We thank the Center for Polymer Studies, Boston University
where part of this work was carried out. We would like to thank
H. E. Stanley and H.  A. Makse for fruitful discussions. This work was
supported by OTKA (T016568, T024004) and MAKA (93b-352).

\bigskip

%\end{multicols}

FIG.\ref{ncoll} Three examples of the collision rules in fluidized
regions.  The left column shows three pre-collision configurations. The
middle and the right column show configurations - together with their
probabilities - after an elastic and an inelastic collision, respectively.
Cases a), b) and c) are examples of a head-on collision between two
particles, with zero, one and two rest particles also present on the site.
In case a), if the collision is elastic, one of two possible configurations
are chosen with equal ($1/2 p_k$) probability. With $1-p_k$ probability the
collision is inelastic and the particles stop. In case b) one scattering
direction is blocked, which is free in example a). In case c) both
directions are blocked and binary collisions between particles on opposite
bonds take place.

FIG.\ref{nfric} An example for the application of the friction rule, when a
moving particle arrives at a bulk site. (The marked particles belong to the
bulk.) With probability $p_\mu$ the particle looses its energy and becomes
part of the bulk. In the opposite case its momentum is conserved and the
node is not considered to belong to the static part any more. 

FIG.\ref{nprop} Illustration of the propagation step. Since two particles
can be present on a bond, binary collisions may take place. Those
particles after an inelastic collision are marked. The figure shows examples
for all possible configurations on the bonds. 

FIG.\ref{calibrate} Dependence of the angle of repose of the pile on the
friction parameter.

FIG.\ref{halfpile} Simulation snapshot of the pile. Note, that each circle
represents a lattice node, which can be occupied by up to seven particles.

FIG.\ref{tdep} The total mass (number of particles) of two piles vs. time.
(One timestep is an interval long enough to contain the avalanches of the
longest lifetime, rather than an update unit). (The friction parameter is
$p_\mu=0.34$). A factor of two in the system size results in significantly
different time sequence.

FIG.\ref{power} The power spectra computed from total mass - time graphs for
two different sizes. (Both curves have been averaged for 25 runs.) The peak
at $f=2.5\  10^{-4}$ ( $\Delta t\approx 4000$) in case of $L=48$ shows the
appearance of a characteristic time scale.

FIG.\ref{tscal} Finite size scaling of lifetime distributions
($p_\mu=0.5$). The best collapse is obtained at the scaling exponents
$\lambda=0.8\pm 0.02$, $\alpha=1.6\pm 0.04$. The exponent of the power law
part of the distribution is $y=1.92\pm0.05$. All distribution curves on
this graph and on the following ones are binned.

FIG.\ref{mscal} Finite size scaling of droplet distributions ($p_\mu=0.5$).
The $\beta=1.4$, $\mu=0.7$ exponents here used for visualisation only,
since it is apparent that a scaling according to the ansatz in the text
cannot be applied.

FIG.\ref{strech} A streched exponential fit $f(M,20)\sim \exp(-c*M^\gamma)$ for
the droplet probability density, in the case of a small pile.
$\gamma=0.34$ is used here, which is equal to the fit used for experimental
data. (The value of $c$ is $2.0\pm 0.2$ in the model, which depends on the
details of the dynamics.)

FIG.\ref{fricdep} The effect the friction coefficient on the droplet 
distribution. The system size is $L=80$. 

FIG.\ref{dissip} Probability density of the dissipated energy in an
avalanche for four system sizes. The exponent of the power-law part of the
curves is $1.63\pm0.06$. 

\eject

\end{document}